\documentclass[12pt]{article}
\usepackage{pqft}
\addtocounter{page}{142}
\begin{document}
\title{On the faster-than-light motions in electrodynamics}
\author{G. A. Kotel'nikov}
\institutes{
RRC Kurchatov Institute, Moscow 123182, Russia}
\maketitle
\abstract{A variant of electrodynamics is constructed in which
the faster-than-light motions are possible.}
\authorrunning{G. A. Kotel'nikov}
\titlerunning{On the faster-than-light motions in electrodynamics}
The existence   of   faster-than-light   motions   (with    velocities
$v>3\cdot10^{10}$  cm/sec)  is  the  subject  of  discussion in modern
physics.
\begin{sloppypar} It  is  in  1946-1948   when   Blokhintsev   already
\cite{paper1} paid attention to the possibility of formulating a field
theory   that   permits   the   propagation    of    faster-than-light
(superluminal)  interactions outside the light cone.  Kirzhnits (1954)
\cite{paper3} showed that a particle possessing  the  tensor  of  mass
${M^i}_k=diag(m_0,m_1,m_1,m_1)$  may  move  with the faster-than-light
velocity if $m_0>m_1$.  Terletsky (1960) \cite{paper4} introduced into
theoretical   physics  the  particles  with  imaginary  masses  moving
faster-than-light. Feinberg (1967) \cite{paper5} named these particles
tachyons   and  described  their  main  properties.  Research  on  the
superluminal tachyon motions opened up additional opportunities  which
were  studied by many authors (hundreds publications),  for example by
Bilaniuk and Sudarshan~(Sb. \cite{paper5}), Kirzhnits and Sazonov (Sb.
\cite{paper5}),  Recami \cite{paper6},  Mignani (Mon.  \cite{paper6}),
Corben (Mon. \cite{paper6}). \end{sloppypar}

The publications are also known in which the violation  of  invariance
of  the  speed  of light is considered \cite{paper7} - \cite{paper13}.
One can note,  for example,  Pauli monograph  \cite{paper7}  with  the
elements of Ritz and Abraham theories;  Logunov lections on Relativity
Theory \cite{paper8};  Glashow paper \cite{paper9} on the violation of
Lorentz-invariance  in  astrophysics;  publications  \cite{paper10}  -
\cite{paper13} considering the violation of invariance of the speed of
light in SR.

Below a  version of the theory permitting faster-than-light motions of
electromagnetic fields and  charged  particles  with  real  masses  is
proposed as the continuation of such investigations.

Let us introduce space-time $R^4$  with the metric
\begin{equation}\label{d}
\displaystyle
\begin{array}{c}
ds^2=({c_0}^2+v^2){dt}^2-{dx}^2-{dy}^2-{dz}^2=                      \\
({c'_0}^2+v'^2)d{t'}^2-d{x'}^2-d{y'}^2-{dz'}^2 - invariant,
\end{array}
\end{equation}
where $t$ is the time,  $x,y,z$ are the spatial variables,  $v$ is the
velocity of a particle being investigated,  $c_0$ is the proper  value
of the speed of light.  Due to homogeneity and isotropy of space,  the
velocity $v$ does not depend on space-time variables.  Let the  proper
value of the speed of light be invariant:
\begin{equation}\label{c}
c_0={c'}_0=3\cdot10^{10} cm/sec.
\end{equation}
As a  result,  the  common time may be introduced on the trajectory of
movement of frame $K'$ and the velocity of light $c$, corresponding to
the  velocity $v$:  $dt=dt'_0 \to c=c_0\sqrt{1+v^2/{c_0}^2}$ \footnote
{In the form of $c'=c(1-\beta^2)^{1/2}$ this formula was  obtained  by
Abraham   before  \cite{paper7}.}.  Let  us  introduce  also  the  new
variables
\begin{equation}\label{nt}
\displaystyle 
x^0=\int\limits_{0}^{t} cd\tau=
c_0\int\limits_{0}^{t}\sqrt{1+{v^2}/{c_0}^2}d\tau, \ \
x^{\alpha}=(x,y,z), \
\alpha=1,2,3,
\end{equation}
and turn  to  Minkowski  space  $M^4$  with  the  metric 
\begin{equation}\label{ds}
ds^2=(dx^0)^2-(dx^1)^2-(dx^2)^2-(dx^3)^2 - invariant.
\end{equation}
The infinitesimal transformations,  retaining the  invariance  of  the
form $ds^2$ are $d{x'}^i={L^i}_kdx^k,  \ i,k=0,1,2,3,$ where ${L^i}_k$
is the matrix  of  Lorentz  group  \cite{paper15}.  The  corresponding
integral transformations are
\begin{equation}
x'^0=\frac{x^0-\beta x^1}{\sqrt{1-\beta^2}}; \
x'^1=\frac{x^1-\beta x^0}{\sqrt{1-\beta^2}}; \ x'^{2,3}=x^{2,3}; \
c'=c\frac{1-\beta u^1}{\sqrt{1-\beta^2}}.
\end{equation}
They belong  to  the  group  of direct product $L_6{\rm X}S_1$,  where
$L_6$ is the Lorentz group,  $S_1$ is the scale transformations  group
$c'=\gamma  c$.  One  can  say  that  these transformations act in the
5-space $V^5=M^4XV^1$,  where $V^1$ is a subspace of the velocities of
light  $c$.  The  relationship  between  the  partial  derivatives  of
variables $(t,x,y,z)$ and $(x^0,x^1,x^2,x^3)$ are as follows:
\begin{equation}
\begin{array}{c}  
\displaystyle      
\frac{\partial}{\partial t}=\frac{\partial
x^0}{\partial t}\frac{\partial}{\partial x^0}+
\sum_{\alpha}\frac{\partial x^{\alpha}}{\partial t}\frac{\partial}
{\partial x^{\alpha}}=c\frac{\partial}{\partial x^0};               \\
\vspace{2mm}
\displaystyle \frac{\partial}{\partial x}=\frac{\partial x^0}{\partial
x}      \frac{\partial}{\partial      x^0}+\sum_{\alpha}\frac{\partial
x^{\alpha}}    {\partial    x}\frac{\partial}{\partial    x^{\alpha}}=
\Big(\int\limits_{0}^{t}\frac{\partial c}{\partial x}d\tau\Big)
\frac{\partial }{\partial x^0}+
\frac{\partial}{\partial   x^1}.                                    
\end{array} 
\end{equation}
The expressions  for  $\partial/\partial  y$ and $\partial/\partial z$
are analogous  to  the  expression  $\partial/\partial  x$.  Below  we
restrict  our  study  by  the case,  when the velocity of light in the
range of interaction does not depend on the space variables. Then
\begin{equation}\label{nab}
\displaystyle
\nabla c(x^0)=0\to \nabla c(t)=0.
\end{equation}
Some features of this study are:                                    \\
1. As in SR, parameter $\beta=V/c$ in the present work is in the range
$0\leq\beta<1$.                                                     \\
2. As in SR, the value $dx^0$ is the exact differential.            \\
3. The time $x^0=\int cdt$ is the functional of $c(t)$ in general case.                               \\
4. The property $\beta$=const is compatible with $V(t), \ c(t)$.

Keeping this in mind,  let us construct a theory in $M^4$ and  reflect
it  on  the space $R^4$.  Following \cite{paper14},  we start from the
expression for integral of action:
\begin{equation}\label{act}
\begin{array}{c}
\displaystyle
S={S}_m+{S}_{mf}+{S}_f=-mc_0\int ds-\frac{e}{c_0}\int A_i dx^i-
\frac{1}{16\pi c_0}\int F_{ik}F^{ik} d^4 x.
\end{array}
\end{equation}
Here $S_m$ is the action for a free particle;  $S_f$ is the action for
a  free electromagnetic field;  $S_{mf}$ is the action for interaction
between a charge $e$ and electromagnetic field;  $m$ is the mass of  a
particle,   $j^i=(\rho,\rho{\bf  u}^{\alpha})$  \cite{paper7}  is  the
4-vector   of   current   density;   \   ${\bf    u}^{\alpha}=    {\bf
v}^{\alpha}/{c}$  is the 3-velocity of a particle.  The meaning of the
other values is standard.  Accordingly to the construction, the action
is Lorentz invariant and does not depend on the velocity of light $c$.
As a result the action is also invariant with respect to the group  of
direct product $L_6{\rm X}S_1$. Lagrangian takes the form:
\begin{equation}\label{Lag}
L=-mc_0\sqrt{1-u^2}+\frac{e}{c_0}({\bf A}\cdot{\bf u}-\phi).
\end{equation}
The generalized  momentum  ${\bf P}$ and generalized energy ${\cal H}$
are:
\begin{equation}\label{mom}
\begin{array}{c}
\vspace{1mm}
\displaystyle
{\bf P}=\frac{\partial  L}{\partial{\bf u}}=\frac{mc_0{\bf u}}{\sqrt
{1-u^2}}+\frac{e}{c_0}{\bf A}={\bf p}+\frac{e}{c_0}{\bf A}=m{\bf v}+
\frac{e}{c_0}{\bf A},                                               \\
\displaystyle
{\cal H}={\bf P}\cdot{\bf u}-L=(mc_0c+e\phi)/c_0.
\end{array}
\end{equation}
Here ${\bf       p}=m{\bf       v}$       is       the       momentum,
$mc_0c=m{c_0}^2(1+v^2/{c_0}^2)^{1/2}={\cal    E}$   is   the   energy,
$T=m{c_0}^2(c/c_0-1)$ is the kinetic energy of a particle.  For a free
particle they are integrals of motion, ${\cal E}$ and ${\bf p}$ may be
united into the 4-momentum $p^i$ (as in SR)
\begin{equation}\label{p}
p^i=mc_0u^i=\Big(\frac{mc_0c}{c_0},mcu^{\alpha}\Big)=
\Big(\frac{{\cal E}}{c_0},m{\bf v}\Big).
\end{equation}
The components of the 4-momentum are related by the expressions:
\begin{equation}\label{dis}
\displaystyle
p_ip^i=\frac{{\cal   E}^2}{{c_0}^2}-{\bf    p}^2={m}^2{c_0}^2;   \ \
{\bf   p}=\frac{{\cal   E}}{c_0c}{\bf   v}; \
{\bf p}=\frac{{\cal E}}{c_0c}{\bf c}, \ if \ m=0, \ {\bf v}={\bf c}.
\end{equation}
One can see from here that the momentum of a particle  with  the  zero
mass  $m=0$  does  not  depend from the particle velocity $v=c$ and is
only determined by  the  particle  energy  in  accordance  with  ${\bf
p}={\bf n}{\cal E}/c_0, \ {\bf n}={\bf c}/c$.

Next we  start  from   the   mechanical   \cite{paper14}   and   field
\cite{paper14, paper15} Lagrange equations:
\begin{equation}\label{Lagr}
\displaystyle
\frac{d}{dx^0}\frac{\partial L}{\partial{\bf u}}-
\frac{\partial L}{\partial{\bf x}}=0; \
\frac{\partial}{\partial x^k}\frac{\partial {\cal L}}
{\partial(\partial
A_i/\partial x^k)}-\frac{\partial {\cal L}}{\partial A_i}=0.
\end{equation}
\begin{sloppypar}\noindent
Here $L$ is the  Lagrangian;  ${\cal  L}=-(1/c_0)A_ij^i-(1/16\pi  c_0)
F_{ik}F^{ik}$    is    the   density   of   the   Lagrange   function;
$\partial(F_{ik}F^{ik})/\partial(\partial A_i/\partial  x^k)=-4F^{ik}$
\cite{paper14}.  As  a  result,  we find the equations of motion for a
charged particle and electromagnetic field:
\end{sloppypar}
\begin{equation}\label{mov}
\begin{array}{c}
\vspace{1mm}
\displaystyle
\frac{d{\bf p}}{dt}=m_0\frac{d{\bf v}}{dt}=\frac{c}{c_0}e{\bf E}+
\frac{e}{c_0}{\bf v}{\rm x}{\bf H};                                 \\
\displaystyle
\frac{d{\cal E}}{dt}=e{\bf E}\cdot{\bf v}\to m_0\frac{dc}{dt}=
\frac{e}{c_0}{\bf v}\cdot{\bf E}.
\end{array}
\end{equation}
\begin{equation}\label{max}
\begin{array}{ll}
\vspace{1mm}
\displaystyle
\nabla{\rm X}{\bf E}+\frac{1}{c}\frac{\partial{\bf H}}{\partial t}=0; &
\displaystyle
\nabla\cdot{\bf E}=4\pi\rho;                                         \\
\displaystyle
\nabla{\rm X}{\bf H}-\frac{1}{c}\frac{\partial{\bf E}}{\partial t}=
4\pi\rho\frac{{\bf v}}{c};                                            &
\nabla\cdot{\bf H}=0,
\end{array}
\end{equation}
\begin{sloppypar} \noindent
where               $c(t)=c_0(1+v^2/{c_0}^2)^{1/2}=c(0)[1+(e/mc_0c(0))
\int\limits_{0}^{t}{\bf v}\cdot{\bf E}d\tau\Big]$, $\nabla c=0$. These
equations,  considered  as  the  whole,  form  the  set  of  nonlinear
equations  of  electrodynamics.   They   permit   the   existence   of
faster-than-light  motion  of a particle with the real mass $m$,  rest
energy ${\cal E}_0=m{c_0}^2$ and velocity
\end{sloppypar}
\begin{equation}\label{v}
v=\sqrt{{\cal E}^2-{m}^2{c_0}^4}/mc_0>c_0,
\end{equation}
if the particle energy is ${\cal E}>\sqrt2 {\cal E}_0$ \cite{paper13}.
For example,  for electron $\sqrt2{\cal E}_0$ is $723~{\rm keV}$,  the
velocity of $1~{\rm GeV}$ electron is $\sim 2000~c_0$.  The ${\cal E}$
and ${\bf p}$ variations with time determine a new dynamics which goes
to Newton one if $v^2/{c_0}^2\ll 1$.

Explained in  the  framework  of  the  constructed  theory  may be the
Michelson  experiment,  Fizeau  experiment,   aberration   of   light,
dilatation of life time of atmospheric $\mu$ - mesons, Doppler effect,
known tests to check independence of  the  speed  of  light  from  the
velocity  of  the  light  source,  Compton  effect,  decay of unstable
particles,  creation of new particles in nuclear  reactions,  possible
faster-than-light  motions of nuclear reaction products.  For example,
in the case of Compton effect \cite{paper16} we find from the  law  of
energy-momentum              conservation:              $\displaystyle
\hbar\omega=\hbar\omega'+m{c_0}^2
\Big[\Big(1+{v^2}/{{c_0}^2}\Big)^{1/2}-1\Big];   \   \   \displaystyle
({\hbar\omega}/{c_0})=({\hbar\omega'}/{c_0})cos\theta+mvcos\alpha,   \
0=({\hbar\omega'}/{c_0})sin\theta-mvsin\alpha.$ \ Here $\hbar\omega, \
\hbar\omega'$ are the energies  of  incident  and  scattered  $\gamma$
quanta,  $m{c_0}^2$  is  the  rest  energy  of electron,  $\alpha$ and
$\theta$ are the  angles  of  scattering  the  electron  and  $\gamma$
quantum   respectively.  As  a  result  the  angular  distribution  of
scattered               $\gamma$               quanta               is
$\omega'=\omega/[1+\hbar\omega(1-cos\theta)/m{c_0}^2]$.   (As   in  SR
\cite{paper16}).  But the velocity of forward-scattered  electron  may
exceed     the     speed     of    light    $c_0$:    ${v_e}(\alpha=0)
\footnote{${v_e}(\alpha=0)=
c_0(\hbar\omega/m{c_0}^2)[1-m{c_0}^2/(2\hbar\omega+m{c_0}^2)]\approx
\hbar\omega/mc_0-0.5c_0$}       =\hbar\omega/mc_0-0.5c_0>c_0$       if
$\hbar\omega>1.5m{c_0}^2$,   which   is  in  accordance  with  formula
(\ref{v}) and  which  differs  from  SR.  The  velocity  of  scattered
$\gamma$  quantum  does  not  depend  on  the  angle  $\theta$  and is
determined by mechanism of scattering (immediately,  or in the act  of
absorption-emission by scattered electron).

As a  result  the  theory has been constructed which is invariant with
respect to the group of direct product $L_6{\rm X}S_1$.  In accordance
with  \cite{paper1}  we  may  assume  that  the proposed theory may be
useful in the field of quantum physics of  extended  particles,  where
the  property  of  elementary  nature  should  not  contradict  to the
existence of the internal structure of a particle.


\small
\begin{thebibliography}{99}\itemsep=0pt
\bibitem{paper1} D. I. Blokhintsev, JETP, \textbf{16}, 480-482 (1946);
\textbf{18}, 566-574 (1948).
\bibitem{paper3} D. A. Kirzhnits, JETP, \textbf{27}, 6-18 (1954).
\bibitem{paper4} Ya.  P. Terletsky, DAN (Russ.), \textbf{133}, 329-332
(1960).
\bibitem{paper5} G.  Feinberg, {\it Einstein Sbornik 1973} (M., Nauka,
1974), pp. 134-177.
\bibitem{paper6} E.~Recami,   {\it   Relativity   Theory    and    its
Generalization.  Astrophysics,  Quanta  and  Relativity  Theory}  (M.,
Izdatelstvo Mir, 1982) pp. 53-128.
\bibitem{paper7} W.  Pauli,  {\it  Theory   of   Relativity}   (M.-L.,
Gostexizdat, 1947) pp. 29, 116.
\bibitem{paper8} A.  A.  Logunov,  {\it  Lections  on  Fundamentals of
Relativity Theory} (Moscow State University, 1982) pp. 25-50.
\bibitem{paper9} S.   L.   Glashow,  Nucl.  Phys.  B  (Proc.  Suppl.),
\textbf{70}, 180-184 (1999).
\bibitem{paper10} P.~M.~Rapier, IRE, \textbf{50}, 229-230 (1962).
\bibitem{paper11} J.~P.~Hsu and L.~Hsu,  Phys.  Lett. A, \textbf{196},
1-6 (1994).
\bibitem{paper12} A.~Chubykalo and  R.~Smirnov-Rueda,  Phys.  Rev.  E,
\textbf{53}, 5373-5381 (1996).
\bibitem{paper13} G.~A.~Kotel'nikov,  J.  of Russian  Laser  Research,
\textbf{22}, 455-474 (2001).
\bibitem{paper14} L.~D.~Landau and E.~M.~Lifshitz.  {\it The Theory of
Field} (M., Fizmatgiz, 1973) pp. 68-71, 93-95, 110.
\bibitem{paper15} N.~N.~Bogoliubov     and     D.~V.~Shirkov.     {\it
Introduction to the Theory of Quantized Fields} (M., Nauka,  1973) pp.
16-17.
\bibitem{paper16} D.  I.  Blokhintsev,  {\it  Fundamentals  of Quantum
Mechanics} (M.-L., Gostexizdat, 1949) pp. 17-18.
\end{thebibliography}
\end{document}